\documentclass[showpacs,preprintnumbers,prd,nofootinbib,floats,amssymb,floatfix]{revtex4}
\usepackage{graphicx}
\usepackage{amsxtra}
\usepackage{hyperref}
\usepackage{amssymb}
\usepackage{amstext}
\usepackage{amsmath}
\usepackage{cleveref}
\usepackage{stackrel}
\usepackage{pdflscape}
\usepackage{nccmath}
\usepackage{blkarray}
\begin{document}
\title{ New canonical  analysis for consistent extension of $\lambda R$ gravity}
\author{Alberto Escalante}  \email{aescalan@ifuap.buap.mx}
\author{P. Fernando Ocaña-Garc{\'i}a}  \email{pfgarcia@ifuap.buap.mx}
 \affiliation{Instituto de F\'isica, Benem\'erita Universidad Aut\'onoma de Puebla. \\ Apartado Postal J-48 72570, Puebla Pue., M\'exico, }

\begin{abstract}
The canonical analysis of the $\lambda R$ model extended with the term due to Blas, Pujolas, and Sibiryakov $[BPS]$ is performed. The analysis is developed for any value of $\lambda$, but particular attention is paid to the point $\lambda=\frac{1}{3}$ because of the closeness with linearized General Relativity [GR]. Then, we add the higher-order conformal term, the so-called Cotton-square term, to study the constraint structure of what constitutes an example of kinetic-conformal Horava’s gravity. At the conformal point, an extra second-class constraint appears; this does not arise at other values of $\lambda$. Then, the Dirac brackets are constructed, and we will observe that the $\lambda R$-Cotton-square model shares the same number of degrees of freedom with linearized $GR$.

\end{abstract}
 \date{\today}
\pacs{98.80.-k,98.80.Cq}
\preprint{}
\maketitle

%Hereby the theory is constructed only with higher spatial derivatives and  avoids the ghost problems that usually arise due to higher time derivatives.
%------------------------------------------------------------------------------------------------------------------------------------------------------.----------------------------------------
%to drive the behaviour in the UV$

\section{Introduction}
Hořava gravity is a higher-order theory that stands out as a serious candidate for generating a complete quantum gravity theory by following the schemes of perturbative quantum field theory \cite{H1,H2,He}. The remarkable features of Hořava gravity are a preferred foliation of spacetime and invariance under the group of diffeomorphisms preserving this structure (FDiff), which conveniently allows us to consider anisotropy. The proposal of breaking the local Lorentz symmetry is intended to avoid the ghosts due to higher-time derivatives \cite{OD}, representing an alternative path to solve the problems of unitarity and renormalizability \cite{Stell, G}.  In fact, the renormalizability of the projectable version has been proven in \cite{Ba}, whereas the quantization of the non-projectable case has been hampered due to the difficulty posed by the presence of second-class constraints \cite{bellorin6,bellorin7,bellorin8}. Remarkably, a proof has recently been presented in which the quantization is performed through the Batalin-Fradkin-Vilkovisky (BFV) formalism, and the renormalization is achieved by using the approach of Barvinsky et al. based on the background field formalism \cite{Renor}.\\ Regarding the  underlying structure, the preservation of the foliation provides an absolute distinction between time and space similar to the Newtonian one, which allows anisotropy  by assuming  a different scaling between space and time according to 
\begin{equation}
    t\rightarrow b^{-z}t,\hspace{1cm}  x^{i}\rightarrow b^{-1}x^{i},
\end{equation}
  where  $z$ is the so-called critical exponent. On this basis, the theory is constructed with a potential containing terms with spatial derivatives of different orders and a kinetic part that employs solely time derivatives of order two. In order to ensure power-counting renormalizability,  at least six-order terms must be considered in $3+1$ dimensions \cite{H1}.\\
Concerning the kinetic part,  a central aspect in setting up the Hořava action is the introduction of the $\lambda$ parameter that determines the separate compatibility of the kinetic terms with FDiff. An outstanding feature due to its dynamic implications is that, in the realm of non-projectability, the kinetic part acquires an anisotropic conformal symmetry at  $\lambda=\frac{1}{3}$ \cite{H1}. The Weyl transformations are anisotropic in the sense that the lapse function scales with a weight different from the one of the spatial metric and the shift vector
\begin{equation}
\Tilde{g}_{ij}=\Omega^{2}g_{ij},\hspace{.5cm}\Tilde{N}=\Omega^{3}N,\hspace{.5cm}\Tilde{N}_{i}=\Omega^{2}N_{i},
\end{equation}
where the dependence of $\Omega=\Omega(x,t)$ is consistent only with non-projectability.  Thus,  by including the extension of the non-projectable potential provided by Blas, Pujol\'as, and Sibiryakov (BPS) \cite{blasf}, we can generate a full anisotropic conformal theory if the potential is chosen to be conformal. Conversely, if the potential is not conformal, we obtain a not-conformal gravitational theory, which is called the kinetic-conformal Hořava theory \cite{bellorin2}.  In both cases, the conformal symmetry of the kinetic part gives rise to a primary constraint that decreases the degrees of freedom of the theory, propagating two as in $GR$. However, this constraint changes from second-class in the kinetic-conformal case to a gauge symmetry associated with infinitesimal conformal transformations in the anisotropic-conformal case   \cite{bellorin2}. An excellent analysis of the dynamics of both versions is performed in \cite{Dy}. \\
On the other hand, the physical feasibility of Hořava's proposal can be assessed by its behavior to low energies, say,  long distances. In this sense, the original non-projectable Hořava gravity leads to the  $\lambda R$ model,   conformed by the $z=1$ compatible terms. It is essential to highlight that although the Einstein-Hilbert action is obtained here identically when $\lambda=1$, i.e., when it is restored the full diffeomorphism symmetry, the compatibility can be achieved regardless  $\lambda$ \cite{bellorin}. Consequently,  the non-projectable case has field equations closer to $GR$. It is worth  mentioning  that implementing a complete canonical analysis has been fundamental to consolidating these results \cite{bellorin, Oca}. Indeed, it is well known that the fundamental aspects of gauge theories can be better handled by employing this formalism \cite{Dirac, Henn}. In a previous work \cite{Oca}, the canonical analysis of perturbative $\lambda R$ gravity was performed by implementing a $3+1$ scheme based on the introduction of an extrinsic curvature type variable. This approach allowed for a closer identification of the constraints, just like in the familiar analysis of perturbative  $GR$ is done \cite{Bar}.\\ 
With all discussed above, in this paper,  by using the framework reported in \cite{Oca}, we perform a perturbative canonical analysis of the $\lambda R$  model extended with its corresponding  $z=1$  $BPS$  term. The complete set of constraints and their classification into first and second-class for any value of $\lambda$ are reported; in the analysis, we consider special attention for $\lambda=\frac{1}{3}$. This extended version gives rise to a prime example of kinetic-conformal Hořava theory at this value. On the other hand,  we also add the square of the Cotton tensor to the potential being a conformal six-order term and the analysis is developed. In this model, we study the structure of the constraints when considering a higher potential in what is regarded as a theory with the soft breaking of conformal symmetry \cite{bellorin2}:  although the Cotton-square term is conformally invariant, the extended $\lambda R$ is not. \\
The paper is organized as follows. In section II, we present the perturbative analysis of the system  $\lambda R$ plus the $BPS$ term for any value of $\lambda$, with an emphasis on the modification to the constraint structure generated at the point $\lambda=\frac{1}{3}$. In section III the $\lambda R$ plus the  Cotton-square term is analyzed. We report the constraint structure, and counting the degrees of freedom is carried out. Finally, the Dirac brackets are constructed for each case. 
\section{The $\lambda R$ gravity plus the BPS term}
As we commented above, Hořava theory is  grounded in the group of  diffeomorphism that preserves the foliation, given by 
\begin{equation}
    t\rightarrow t'(t),\hspace{2cm}x^{i}\rightarrow x'^{i}(\Vec{x},t),
\end{equation}
 in coordinates adapted to the foliation. The analysis of  the dynamics concerning this particular gauge group has been of great importance at the classical level. Its very structure suggests the presence of a strongly coupled additional degree of freedom to the two of GR \cite{c1}. Although it has been possible to find useful cosmological applications for this extra mode \cite{Noemi, ParkD, ParkD2}, initially, this put into debate the consistency of Hořava theory, which at the IR regime would differ from $GR$ and its well-tested predictions. Remarkably, the $\lambda R$ theory has been shown to be fully consistent with  $GR$ despite the reduced symmetry group \cite{bellorin, Oca}. Here,  the extra scalar mode that occurs in the full action is suppressed by the emergent constraints structure, thus only propagating two degrees of freedom.\\
On the other hand, the extended $\lambda R$ model is the lowest-order effective action, up to second order in derivatives, of the complete Hořava theory;  it is constructed including the $z=1$ $BPS$  term that is symmetry-compatible and depends on the FDiff-covariant vector $a_{i}=\partial_{i}\mathrm{ln}\hspace{0.5mm}N$, resulting in the following second-order action written in the Arnowitt-Deser-Misner formalism (ADM) \cite{ADM}.
\begin{equation}\label{full}
    S=\int dt d^{3}x\sqrt{g}N\left(K_{ij}K^{ij}-\lambda K^{2}+R+\alpha a_{i}a^{i}\right),
\end{equation}
where $N$ is the lapse function, and its dependence on $x$ and $t$ characterizes the non-projectability,  $R$ is the spatial Ricci scalar, $K_{ij}=\frac{1}{2N} \left(\dot{g}_{ij}-2\nabla_{(i}N_{j)}\right)$ is the extrinsic curvature and $g_{ij}$ is the Riemannian spatial metric.  \\
 Our canonical analysis will focus on the action (\ref{full}) but it will be carried out in the perturbative sector. Although the action is originally written regarding ADM variables, we will use a different analysis method. Namely, we will use the perturbative 3+1 formalism, which is also compatible with the preferred time direction defined by FDiff, and it is helpful to economize the analysis at the perturbative level \cite{Oca, Fh, OC}. The implementation of this formalism starts by considering the well-known Fierz-Pauli Lagrangian for massless particles of spin 2 \cite{fierz}. 
\begin{equation}\label{FP}
    \begin{split}
        \mathcal{L}_{FP}=&\frac{1}{4}\Dot{h}_{ij}\Dot{h}^{ij}-\Dot{h}^{ij}\partial_{i}h_{0j}-\dot{h}_{j}^{j}\partial_{i}h^{0i}-\frac{1}{4}(\dot{h}_{i}^{i})^{2}-\frac{1}{2}\partial_{i}h_{0j}\partial^{i}h^{0j}+\frac{1}{2}\partial^{i}h^{j0}\partial_{j}h_{i0}+\frac{1}{2}\partial_{i}h_{00}\partial_{j}h^{ij}\\&\hspace{4mm}-\frac{1}{2}\partial_{i}h_{k}^{k}\partial_{j}h^{ij}-\frac{1}{2}\partial_{i}h_{00}\partial^{i}h_{k}^{k}+\frac{1}{4}\partial_{i}h_{j}^{j}\partial^{i}h_{k}^{k}+\frac{1}{2}\partial^{i}h^{jk}\partial_{j}h_{ik}-\frac{1}{4}\partial_{i}h_{jk}\partial^{i}h^{jk}.
    \end{split}
\end{equation}
This action describes linearized gravity  on a Minkowski background  and it is written in its $3+1$ form; the perturbation is given by $g_{\ mu\nu}=\eta_{\mu\nu}+h_{\mu\nu}$ with $\eta_{\mu\nu}=\mathrm{diag}(-1,+1,+1,+1)$.  The complete compatibility with FDiff is established by introducing an extrinsic curvature type variable given by $ K_{ij}=\frac{1}{2}\left(\Dot{h}_{ij}-\partial_{i}h_{0j}-\partial_{j}h_{0i}\right)$, and expressing the kinetic part of $\mathcal{L}_{FP}$  in a new fashion. In fact, the action will be written in a Horava-like form, and it will be in agreement with the action (\ref{full}). Thus, introducing $K_{ij}$ and  adding the linearized  $BPS$ term into $\mathcal{L}_{FP}$, we obtain 
\begin{equation}\label{Lambda1}
    \mathcal{L}=G^{ijkl}K_{ij}K_{kl}-\frac{1}{2}h^{00}R-\frac{1}{2}h^{ij}\left(R_{ij}-\frac{1}{2}\delta_{ij}R\right)+\alpha \partial_{i}h_{00}\partial^{i}h^{00},
\end{equation}
where 
\begin{equation}\label{Lambda2}
    G^{ijkl}=\frac{1}{2}\left(\delta^{ik}\delta^{jl}+\delta^{il}\delta^{jk}\right)-\lambda\delta^{ij}\delta^{kl},
\end{equation}
and the spatial part has been condensed by using
\begin{equation}
    \begin{split}
 R_{ij}&=\frac{1}{2}\left(\partial_{k}\partial_{i}h^{k}_{j}-\partial^{k}\partial_{k}h_{ij}-\partial_{j}\partial_{i}h^{k}_{k}+\partial_{j}\partial^{k}h_{ik}\right),\\  R&=\partial_{i}\partial_{j}h^{ij}-\nabla^{2}h_{i}^{i}.
    \end{split}
\end{equation}
The $\lambda$ parameter occurs in (\ref{Lambda1})  using (\ref{Lambda2}), and the introduction of $G^{ijkl}$ is relevant to simplify  the calculation of the canonical momenta as we will see below.
\subsection{Canonical analysis for $\lambda\neq\frac{1}{3}$}
Our analysis is based on the formalism developed by Dirac-Bergamann for singular systems \cite{Henn}, thus  we start by calculating  the canonical momenta. Since the expression (\ref{Lambda1}) does not depend on the velocities $\dot{h}_{00}$ and $\dot{h}_{0i}$, its conjugate momenta, $\pi^{00}$ and $\pi^{0i}$ respectively, will be primary constraints. On the other hand, the canonical momenta conjugate to $h_{ij}$ is given by
\begin{equation}\label{cons1}
\hspace{1.3cm}\pi^{ij}=\frac{\partial\mathcal{L}}{\partial \dot{h}_{ij}}=G^{ijkl}K_{kl}.
\end{equation}
We use this to obtain an expression for the velocities $\dot{h}_{ij}$ in terms of its conjugated momenta that will be employed to perform the Legendre transformation.
\begin{equation}\label{hvel}\dot{h}_{ij}=2\mathcal{G}_{ijkl}\pi^{kl}+\partial_{i}h_{j0}+\partial_{j}h_{i0},
\end{equation}
where $\mathcal{G}_{ijkl}=\frac{1}{2}(\delta_{ik}\delta_{jl}+\delta_{il}\delta_{jk})+\frac{\lambda}{1-3\lambda}\delta_{ij}\delta_{kl}$ is the inverse of $G^{ijkl}$ and is defined only  for $\lambda\neq \frac{1}{3}$. As we will see below,  due to the impossibility of inverting (\ref{cons1}) in the case $\lambda= \frac{1}{3}$,   an additional primary constraint will emerge: the trace of the canonical momenta $\pi^{ij}$. This constraint is the generator of the infinitesimal conformal transformations and is a gauge symmetry only for the anisotropic conformal Hořava gravity \cite{Dy}.  \\
With these ingredients, we  construct the primary Hamiltonian given by 
\begin{equation}
    H=\mathcal{G}_{ijkl}\pi^{kl}\pi^{ij}-2h_{j0}\partial_{i}\pi^{ij}+\frac{1}{2}h^{00}R+\frac{1}{2}h^{ij}\left(R_{ij}-\frac{1}{2}\delta_{ij}R\right)-\alpha \partial_{i}h_{00}\partial^{i}h^{00}+u\pi^{00}+u_{i}\pi^{0i},
\end{equation}
where $u$ and $u_{i}$ are the Lagrange multipliers enforcing the primary constraints $\pi^{00}\approx0$ and $\pi^{0i}\approx0$ respectively. Now, by introducing the fundamental  Poisson-bracket relations  $ \left\lbrace h_{ij}(x),\pi^{kl}(y)\right\rbrace=\delta^{k}_{(i}\delta^{l}_{j)}\delta^{3}(x-y)$, we explore the consistency of the primary constraints, i.e., its preservation in time
\begin{eqnarray}\label{pri2}
  \nonumber 
      \mathcal{H}&:&\bigl\{ \pi^{00},\int d^{3}x\hspace{1mm}H \bigr\}=\frac{1}{2}R+2\alpha\nabla^{2}h_{00}\approx0, \nonumber \\
      \mathcal{H}^{i}&: &\bigl\{ \pi^{0i},\int d^{3}x\hspace{1mm}H \bigr\}=\partial_{j}\pi^{ji}\approx0,
\end{eqnarray}
 thus, we obtain four secondary constraints: $\mathcal{H}\approx0$ is known as the Hamiltonian constraint, and $\mathcal{H}^{i}\approx0$ is the so-called momentum constraint, which is a first-class constraint generating spatial diffeomorphisms. Hence,  $h_{j0}$ can be regarded as the Lagrange multipliers associated with this first-class constraint.\\
The consistency condition on $\mathcal{H}^{i}$ is identically satisfied, while for  $\mathcal{H}$  leads to an equation involving the multiplier $u$, then the generation of constraints ends. We have obtained a set of eight constraints, $(\pi^{00},  \pi^{0i}, \mathcal{H}^{i}, \mathcal{H})$, which, following the scheme, they need to be classified into first-class and second-class constraints. The second-class has at least one non-zero Poisson bracket.  In this case 
\begin{eqnarray} \nonumber\left\lbrace\mathcal{H},\pi^{00}\right\rbrace&:=&2\alpha\nabla^{2}\delta^{3}(x-y),
\end{eqnarray}
is the only non-null Poisson bracket. Thus there are  2 second-class constraints
\begin{eqnarray}\label{sec-class}
  \nonumber  
\chi_{1}&:&\frac{1}{2}R+2\alpha\nabla^{2}h_{00}\approx0, \nonumber \\ \chi_{2}&:&\pi^{00}
\approx0.
\end{eqnarray}
which are the vanishing of the momentum conjugated to $h_{00}$ and the  analogous to the so-called Hamiltonian constraint in linearized $GR$ \cite{Bar}. On the other hand, we obtain the  following  6  first-class constraints
\begin{eqnarray}
  \nonumber 
\Gamma_{1}^{i}&: &\pi^{0i}\approx0,\nonumber\\
\Gamma_{2}^{i}&: &\partial_{j}\pi^{ji}\approx0,
\end{eqnarray}
which are the generators of gauge symmetries. We highlight that there are two degrees of freedom in the perturbative $\lambda R$ gravity at $\lambda\neq\frac{1}{3}$ \cite{Oca}. In fact,  the consistency condition on the Hamiltonian constraint leads to the second-class $\pi=0$, and the evolution of $\pi$ yields another second-class constraint. These two additional second-class constraints contribute to obtaining two degrees of freedom. However, in this extended model,  the counting of degrees of freedom yields 
\begin{equation}
    DOF=\frac{1}{2}(\mathrm{canonical\hspace{1mm}var.}-2(\mathrm{first\hspace{1mm}class\hspace{1mm}c.})-\mathrm{second\hspace{1mm}class\hspace{1mm}c.})=\frac{1}{2}(20-2-2*6)=3,
\end{equation}
one more than linearized $\lambda R$ gravity. Hence,  outside the conformal point, the extended $\lambda R$ model, in this sense, is not equivalent to linearized $GR$. The relevance of adding the $BPS$ extension is related to the behavior of this  additional mode, giving it a description that goes from a first-order  to a second-order equation. That is, turning it into an even mode  \cite{blasf}.\\
On the other hand, since the second-class constraints are not gauge generators, we can  remove them directly by introducing the Dirac brackets
\begin{equation}\label{Dirac}
    \left\lbrace A, B\right\rbrace_{D}= \left\lbrace A, B\right\rbrace-\int dudv\left\lbrace A, \chi_{a}(u)\right\rbrace C^{ab}\left\lbrace \chi_{b}(v), B\right\rbrace,
\end{equation}
where $C^{ab}$ is the inverse of $C_{ab}=\left\lbrace \chi_{a}, \chi_{b}\right\rbrace$ \cite{Henn}. In this way all the dynamical equations of the theory are expressed in terms of (\ref{Dirac}). In our case, by considering the second-class set (\ref{sec-class}), we get the matrix 
\begin{equation}\label{matrix1}
\makeatletter\setlength\BA@colsep{7pt}\makeatother
\arraycolsep=1.4pt\def\arraystretch{1}
   C_{ab}= \begin{blockarray}{ccccc}
        & \chi_{1} & \chi_{2}\\
      \begin{block}{c(cccc)}
        \chi_{1} & 0 & 2\alpha\nabla^{2}\\
        \chi_{2} & -2\alpha\nabla^{2}& 0  \\
      \end{block}
    \end{blockarray}\hspace{1mm}\delta^{3}(x-y).
\end{equation}
As mentioned before, for the calculation of the Dirac brackets we employ its inverse matrix given by 
\begin{equation}\label{matrix2}
\makeatletter\setlength\BA@colsep{7pt}\makeatother
\arraycolsep=1.4pt\def\arraystretch{1}
   C^{ab}=
    \begin{blockarray}{ccccc}
        & \chi_{1} & \chi_{2} \\
      \begin{block}{c(cccc)}
        \chi_{1} &  0 & -1 \\
        \chi_{2} & 1 & 0 \\
      \end{block}
    \end{blockarray}\hspace{1mm}\frac{1}{2\alpha\nabla^{2}}\delta^{3}(x-y).
\end{equation}
Due to the canonical variables involved in the set (\ref{sec-class}),  we observe a change only for brackets related to $h_{00}$. Namely, the fundamental bracket $\left\lbrace h_{00},\pi^{00}\right\rbrace=\delta^{3}(x-y)$ changes to 
\begin{equation}\label{h}
    \left\lbrace h_{00},\pi^{00}\right\rbrace_{D}=0,
\end{equation}
since $\pi^{00}\approx0$ is second-class, and the otherwise null Poisson bracket between $h_{00}$ and $\pi^{ij}$ becomes 
\begin{equation}\label{h2}
    \left\lbrace h_{00},\pi^{ij}\right\rbrace_{D}=\frac{1}{4\alpha\nabla^{2}}\left(\partial^{i}\partial^{j}-\delta^{ij}\nabla^{2}\right)\delta^{3}(x-y).
\end{equation}
 This bracket is associated with the dynamics of the third degree of freedom and is not present in the non-extended model reported in \cite{Oca}. Furthermore, we observe that $\alpha$ can not be zero; this indicates that the field $h_{00}$ is strongly coupled.  \\
On the other side, one of the most important aspects related to the consistency of lowest-order effective Hořava theory is the existence of a solution for the lapse function $N$ (see (\ref{full}))  since it is a dynamical variable which is expected to be fixed by the Hamiltonian constraint. This is, by performing the non-perturbative canonical analysis of  (\ref{full}), the following Hamiltonian constraint is obtained 
\begin{equation}
\label{nonper}
\left( 4\alpha \nabla^2 -R + \mathcal{G}_{ijkl} \frac{\pi^{ij}\pi^{kl}}{g} \right)\sqrt{N}\approx0, 
\end{equation}
that equation becomes relevant because any source of indetermination on $N$ could either lead to inconsistencies of the theory (\ref{full}) or to reinterpret the Hamiltonian constraint as a condition for another variable as was claimed in \cite{bellorin, Hen2, bellorin3}. At the non-perturbative level, the Hamiltonian constraint (\ref{nonper})  is a second-order elliptic PDE for $N$ totally compatible with the standard (flat) asymptotic behavior of all gravitational variables. In fact, if it is taken $\sqrt{N}=1 + n$, and $g_{ij}= \delta_{ij}+ h_{ij}$,  (\ref{nonper}) is reduced to 
\begin{equation}
\label{nonperl}
4\alpha\nabla^2n= R, 
\end{equation}
this is the equivalent Hamiltonian constraint found in (\ref{sec-class}) using our approach. In fact,  we can identify that $h_{00}$ in our formalism is equivalent to the perturbation $n$. The equation (\ref{sec-class}) is a  Poisson equation that can be solved for $h_{00}$ under appropriate boundary conditions; this ensures that the solution for $h_{00}$ exists, and is unique, at least in the sense of distributions. In this manner, our approach complete the results found in the literature. 

\subsection{Canonical analysis for $\lambda=\frac{1}{3}$}

At the kinetic conformal point, the canonical momenta change; thus, in addition to the primary constraints found in the previous section,  one more will arise from the definition of the canonical momenta conjugate to $h_{ij}$, this is 
\begin{equation}\label{m22}
\pi^{ij}=\frac{\partial\mathcal{L}}{\partial \dot{h}_{ij}}=K^{ij}-\frac{1}{3}\delta^{ij}K.
\end{equation}  
The new constraint is given by $\pi\equiv\delta_{ij}\pi^{ij}=0$. This constraint must be of second-class because conformal gauge transformations are not a gauge symmetry of the theory.  Hence, introducing this constraint, the primary Hamiltonian takes the form
\begin{equation}
        H'=\pi^{ij}\pi_{ij}-2\partial_{i}\pi^{ij}h_{0j} +\frac{1}{2}h^{00}R+\frac{1}{2}h^{ij}\left(R_{ij}-\frac{1}{2}\delta_{ij}R\right)-\alpha \partial_{i}h_{00}\partial^{i}h^{00}+u\pi^{00}+u_{i}\pi^{0i}+v\pi,
\end{equation}
where the primary constraints are identified as  $\pi^{00}\approx0$, $\pi^{0i}\approx0$ and $\pi\approx0$,  and $u$, $u_{i}$ and $v$ are their respective Lagrange multipliers.
From consistency on the above primary constraints, the following secondary constraints arise
\begin{eqnarray}\label{sec1}
  \nonumber 
      \mathcal{H}&:&\frac{1}{2}R+2\alpha\nabla^{2}h_{00}\approx0, \nonumber \\
      \mathcal{H}^{i}&: &\partial_{j}\pi^{ji}\approx0,\nonumber\\
      \gamma&:&\nabla^{2}h^{00}+\frac{1}{2}R
      \approx0,
\end{eqnarray}
 evolution of these secondary constraints does not generate any new constraints. On the other hand,  we notice that the sole difference between $\mathcal{H}\approx0$ and $\gamma\approx0$ is given by the $\alpha$ parameter. Thus, if $\alpha\neq\frac{1}{2}$, these constraints are independent and resolvable to produce the following set of independent constraints \footnote{If $\alpha=\frac{1}{2}$ the set of constraints would be $\pi=\pi^{00}=\mathcal{H}\approx0$,
which is a inconsistent odd second-class  set.}. 
\begin{eqnarray}
\label{conpri}
  \nonumber 
      &&\pi^{00}\approx0, \nonumber \\
      &&\pi^{0i}\approx0,\nonumber\\
      &&\pi\approx0,\nonumber\\
      &&\mathcal{H}^{i}\approx0,\nonumber\\
     &&\frac{1}{2}R \approx0,\nonumber\\
      &&2\alpha\nabla^{2}h^{00}\approx0.
\end{eqnarray}
Now, to perform the classification of these constraints, let's calculate the nonzero Poisson brackets between them, this is 
\begin{equation} 
\begin{split}
    \left\lbrace2\alpha\nabla^{2}h^{00},\pi^{00}\right\rbrace&=2\alpha\nabla^{2}\delta^{3}(x-y),\\
    \left\lbrace\frac{1}{2}R,\pi\right\rbrace&=-\nabla^{2}\delta^{3}(x-y).
\end{split}
\end{equation}
Thus we obtain  the following six first-class constraints given by 
\begin{eqnarray}\label{fc1/3}
  \nonumber 
      \Gamma_{1}^{i}&: &\pi^{0i}\approx0,\nonumber\\
      \Gamma_{2}^{i}&: &\partial_{j}\pi^{ji}\approx0,
\end{eqnarray}
 which are the same as in the previous case, and the following four  second-class constraints 
\begin{eqnarray}\label{sc}
  \nonumber  
      \chi_{1}&:&R\approx0, \nonumber \\
       \chi_{2}&:&\pi\approx0,\nonumber\\
       \chi_{3}&:&\pi^{00}\approx0, \nonumber \\
       \chi_{4}&:&\nabla^{2}h^{00}\approx0.
\end{eqnarray}
The difference with the case $\lambda \neq \frac{1}{3}$  is the presence of two additional second-class constraints that modify the dynamics of the theory, which now propagates two degrees of freedom just like in linearized $GR$ \cite{Bar}. In this case, the first and second-class constraints set corresponds to the sets obtained in the non-extended model reported in \cite{Oca}. By fixing the gauge via the Coulomb gauge $\partial_{i}h^{ij}\approx0$, and $h^{0i}\approx0$,  the following non-zero Dirac brackets are obtained 
\begin{equation}\label{secondfull}
\begin{split}
     \left\lbrace h_{ij},\pi^{lm}\right\rbrace_{D}&=\frac{1}{2}\left(\delta_{i}^{l}\delta_{j}^{m}+\delta_{i}^{m}\delta_{j}^{l}\right)\delta^{3}(x-y)-\frac{1}{2\nabla^{2}}\left(\delta_{i}^{m}\partial_{j}\partial^{l}+\delta_{i}^{l}\partial_{j}\partial^{m}+\delta_{j}^{m}\partial_{i}\partial^{l}+\delta_{j}^{l}\partial_{i}\partial^{m}\right)\delta^{3}(x-y)\\
     &-\frac{1}{2}\delta_{ij}\delta^{lm}\delta^{3}(x-y)+\frac{1}{2\nabla^{2}}\left(\delta_{ij}\partial^{l}\partial^{m}+\delta^{lm}\partial_{i}\partial_{j}\right)\delta^{3}(x-y)+\frac{1}{2}\frac{\partial_{i}\partial_{j}\partial^{l}\partial^{m}}{\nabla^{4}}\delta^{3}(x-y),
\end{split}
\end{equation} 
these Dirac's brackets are the same as those reported for linearized $GR$ \cite{Bar} and for $\lambda R$ gravity \cite{Oca}. It is worth commenting that these brackets are $\alpha$ independent; thus, the propagators between the fields are well defined. From the propagators, we will se that the theory at the critical point propagates two massless degrees of freedom.     
\section{Linearized $\lambda R$ gravity plus a Cotton-square term}
Now we consider a Hořava theory with soft breaking of conformal symmetry \cite{bellorin2}. This is composed by the  previous extended $\lambda R$ model that is not conformal and a square term of the Cotton tensor that is conformally invariant. The Cotton term that we shall  add is $C_{ij}C^{ij}$, where 

\begin{equation}
    C^{ij}=\epsilon^{ikl}\nabla_{k}\left(R^{j}_{l}-\frac{1}{4}R\delta^{j}_{l}\right).
\end{equation}
On a Minkowski background, the linearized  Lagrangian now is written as 
\begin{equation}\label{Lambda}
    \begin{split}
        \mathcal{L}=&G^{ijkl}K_{ij}K_{kl}-\frac{1}{2}h^{00}R-\frac{1}{2}h^{ij}\left(R_{ij}-\frac{1}{2}\delta_{ij}R\right)+\alpha \partial_{i}h_{00}\partial^{i}h^{00}-w\partial_{i}R^{j}_{k}\partial^{i}R_{j}^{k}\\&+w\partial^{i}R_{i}^{j}\partial_{k}R^{k}_{j}+\frac{3w}{8}\partial_{i}R\partial^{i}R+\frac{w}{2}\partial_{i}R^{i}_{j}\partial^{j}R,
     \end{split}
\end{equation}
where $w$ is an arbitrary constant. Now we proceed to perform the canonical analysis. 
\subsection{Canonical analysis for $\lambda\neq\frac{1}{3}$}

Since the primary constraints depend only on the kinetic part, these are the same as the respective previous case.  Thus, the primary Hamiltonian takes the form
\begin{equation}
    \begin{split}
H=&\mathcal{G}_{ijkl}\pi^{kl}\pi^{ij}-2h_{j0}\partial_{i}\pi^{ij}+\frac{1}{2}h^{00}R+\frac{1}{2}h^{ij}\left(R_{ij}-\frac{1}{2}\delta_{ij}R\right)-\alpha \partial_{i}h_{00}\partial^{i}h^{00}\\&
+w\partial_{i}R^{j}_{k}\partial^{i}R_{j}^{k}-w\partial^{i}R_{i}^{j}\partial_{k}R^{k}_{j}-\frac{3w}{8}\partial_{i}R\partial^{i}R-\frac{w}{2}\partial_{i}R^{i}_{j}\partial^{j}R
+u\pi^{00}+u_{i}\pi^{0i}.
    \end{split}
\end{equation}
Similarly, since the high-order potential does not involve $h_{00}$ or $h_{0i}$, the consistency of primary constraints results in the following secondary constraints
\begin{eqnarray}\label{pri3}
  \nonumber 
      \mathcal{H}&:&\frac{1}{2}R+2\alpha\nabla^{2}h_{00}\approx0, \nonumber \\
      \mathcal{H}^{i}&: &\partial_{j}\pi^{ji}\approx0.
\end{eqnarray}
The preservation in time of these constraints does not lead us to more constraints. Note that, at the perturbative level, the added potential  does not affect the sets of first and second-class constraints.  Thus, there are  3 degrees of freedom and the  Dirac brackets are those found in (\ref{h}) and (\ref{h2}).  The same would be true for any higher-order potential that does not involve terms that depend on $h_{00}$, such as the BPS terms.

\subsection{Canonical analysis for $\lambda=\frac{1}{3}$}

Following the same above consideration about the primary constraints, now the primary Hamiltonian takes the form
\begin{equation}
    \begin{split}
        H=&
        \pi^{ij}\pi_{ij}-2\partial_{i}\pi^{ij}h_{0j} +\frac{1}{2}h^{00}R+\frac{1}{2}h^{ij}\left(R_{ij}-\frac{1}{2}\delta_{ij}R\right)-\alpha \partial_{i}h_{00}\partial^{i}h^{00}\\&+w\partial_{i}R^{j}_{k}\partial^{i}R_{j}^{k}-w\partial^{i}R_{i}^{j}\partial_{k}R^{k}_{j}-\frac{3w}{8}\partial_{i}R\partial^{i}R-\frac{w}{2}\partial_{i}R^{i}_{j}\partial^{j}R+u\pi^{00}+u_{i}\pi^{0i}+v\pi.
    \end{split}
\end{equation}
In addition to the set (\ref{pri3}), the presence of $\pi$ adds a secondary constraint whose structure is determined by the dependence on $h_{ij}$ of the high-order terms, so that it differs from its counterpart in (\ref{sec1}). Thus, from consistency of $\pi$ we obtain the following constraint 
\begin{equation}
      \gamma:\nabla^{2}h^{00}+\frac{1}{2}R+2w\nabla^{2}\partial_{i}\partial_{j}R^{ij}+\frac{1}{2}w\nabla^{2}\nabla^{2}R\approx0.
\end{equation}

The presence of $\nabla^{2}h^{00}$ in $\mathcal{H}$ and $\gamma$ leads to expressions containing the Lagrange multiplier $u$ when the consistency condition is applied, and as in the previous cases, the preservation of $\mathcal{H}^{i}$ is identically satisfied. Thus we have obtained the complete set of constraints. To make the separation into first and second-class, let us note that the non-zero Poisson brackets between them are
\begin{equation}\label{par}
\begin{split}
    \left\lbrace\gamma,\pi^{00}\right\rbrace&=\nabla^{2}\delta^{3}(x-y),\\
\left\lbrace\pi,\gamma\right\rbrace&=(\nabla^{2}+w\nabla^{2}\nabla^{2}\nabla^{2})\delta^{3}(x-y),\\\left\lbrace\mathcal{H},\pi\right\rbrace&=-\nabla^{2}\delta^{3}(x-y),
\\\left\lbrace\mathcal{H},\pi^{00}\right\rbrace&=2\alpha\nabla^{2}\delta^{3}(x-y).
\end{split}
\end{equation}
Thus, we obtain  six first-class constraints given by 
\begin{eqnarray}\label{fc1/3}
  \nonumber 
      \Gamma_{1}^{i}&: &\pi^{0i}\approx0,\nonumber\\
      \Gamma_{2}^{i}&: &\partial_{j}\pi^{ji}\approx0,
\end{eqnarray}
and the following four  second-class constraints 
\begin{eqnarray}\label{sc2}
  \nonumber  
      \chi_{1}&:&\frac{1}{2}R+2\alpha\nabla^{2}h_{00}\approx0, \nonumber \\
       \chi_{2}&:&\pi\approx0,\nonumber\\
       \chi_{3}&:&\pi^{00}\approx0, \nonumber \\
       \chi_{4}&:&\nabla^{2}h^{00}+\frac{1}{2}R+2w\nabla^{2}\partial_{i}\partial_{j}R^{ij}+\frac{1}{2}w\nabla^{2}\nabla^{2}R\approx0.
\end{eqnarray}
Although there is an obvious modification in the second-class constraints compared to (\ref{sc}), the gauge symmetries associated with the first-class constraints prevail, as well as the propagation of two degrees of freedom. We will now calculate the Dirac brackets that arise from the set (\ref{sc2}). The matrix of the Poisson brackets between the second-class constraints is 
\begin{equation}\label{matrix3}
\makeatletter\setlength\BA@colsep{7pt}\makeatother
\arraycolsep=1.4pt\def\arraystretch{0.85}
   C_{ab}=
    \begin{blockarray}{ccccc}
        & \chi_{1} & \chi_{2} & \chi_{3} & \chi_{4} \\
      \begin{block}{c(cccc)}
        \chi_{1} & 0 & -\nabla^{2} & 2\alpha\nabla^{2} & 0  \\
        \chi_{2} & \nabla^{2} & 0 & 0 & \nabla^{2}+w\nabla^{2}\nabla^{2}\nabla^{2}  \\
        \chi_{3} &-2\alpha\nabla^{2} & 0 & 0 & -\nabla^{2} \\
        \chi_{4} & 0 & -\nabla^{2}-w\nabla^{2}\nabla^{2}\nabla^{2} &\nabla^{2} & 0  \\
      \end{block}
    \end{blockarray}\hspace{1mm}\delta^{3}(x-y),
\end{equation}
and its inverse is given by
\begin{equation}\label{matrix4}
\makeatletter\setlength\BA@colsep{7pt}\makeatother
\arraycolsep=1.4pt\def\arraystretch{0.85}
   C^{ab}=
    \begin{blockarray}{ccccc}
        & \chi_{1} & \chi_{2} & \chi_{3} & \chi_{4} \\
      \begin{block}{c(cccc)}
        \chi_{1} & 0 & 1 & 1+w\nabla^{4} & 0  \\
        \chi_{2} & -1 & 0 & 0 & 2\alpha  \\
        \chi_{3} &-1-w\nabla^{4} & 0 & 0 & 1 \\
        \chi_{4} & 0 & -2\alpha &-1 & 0  \\
      \end{block}
    \end{blockarray}\hspace{1mm}\frac{1}{(1-2\alpha(1+w\nabla^{4})) \nabla^{2}}\delta^{3}(x-y).
\end{equation}
The Dirac brackets that can be built with this matrix are shown below
\begin{equation}
    \left\lbrace h_{00},\pi^{00}\right\rbrace_{D}=0,
\end{equation}
\begin{equation}
    \left\lbrace h_{ij},\pi^{00}\right\rbrace_{D}=0,
\end{equation}
\begin{equation}
\begin{split}
     \left\lbrace h_{ij},\pi^{lm}\right\rbrace_{D}=&\frac{1}{2}\left(\delta_{i}^{l}\delta_{j}^{m}+\delta_{i}^{m}\delta_{j}^{l}\right)\delta^{3}(x-y)+\frac{\delta_{ij}}{2\Xi}\left(\partial^{l}\partial^{m}-\delta^{lm}\nabla^{2}\right)\delta^{3}(x-y)\\&-\frac{\alpha\delta_{ij}}{\Xi}\left(\partial^{l}\partial^{m}-\delta^{lm}\nabla^{2}\right)\left(1+3w\nabla^{4}\right).
\end{split}
\end{equation}
where $\Xi=(1-2\alpha(1+w\nabla^{4})) \nabla^{2}$. Now, to observe the IR effective action, we take $w \rightarrow 0$, and the Dirac brackets are reduced to 
\begin{equation}
\label{bra1}
 \left\lbrace h_{ij},\pi^{lm}\right\rbrace_{D}=\frac{1}{2}\left(\delta_{i}^{l}\delta_{j}^{m}+\delta_{i}^{m}\delta_{j}^{l}-\delta_{ij}\delta^{lm}\right)\delta^{3}(x-y)+ \frac{\delta_{ij}\partial^l \partial^m }{2 \nabla^2}\delta^{3}(x-y). 
\end{equation}
It is worth mentioning that in \cite{gh}, a different higher-order model was studied; it was constructed in terms of linearized $ADM$ variables and considered the dynamical part and a quadratic term $R_{ij}R^{ij}$, this theory propagates two degrees of freedom, and its fundamental brackets were given by (\ref{bra1}). Furthermore, the first class constraints remain  and we can fix the gauge. In fact, by fixing the gauge the following constraints arrive  
\begin{eqnarray}\label{scfix}
  \nonumber  
      \chi_{1}&:&\frac{1}{2}R+2\alpha\nabla^{2}h_{00}\approx0, \nonumber \\
       \chi_{2}&:&\pi\approx0,\nonumber\\
       \chi_{3}&:&\pi^{00}\approx0, \nonumber \\
       \chi_{4}&:&\nabla^{2}h^{00}+\frac{1}{2}R+2w\nabla^{2}\partial_{i}\partial_{j}R^{ij}+\frac{1}{2}w\nabla^{2}\nabla^{2}R\approx0, \nonumber
       \\
       \chi_{5}&:&\pi^{0i}\approx0, \nonumber
       \\
       \chi_{6}&:&h^{0i}\approx0, \nonumber
       \\
       \chi_{7}&:&\partial_{j}\pi^{ji}\approx0, \nonumber
       \\
       \chi_{8}&:&\partial_{j}h^{ji}\approx0.
\end{eqnarray}
The matrix whose entries are the Poisson brackets between  these constraints is given by 
 \begin{equation}
 \begin{aligned}[b]
   \mathbf{\alpha} &= 
   \begin{blockarray}{cccccccccccc}
       & \chi_{1} & \chi_{2} & \chi_{3}& \chi_{4} & \chi_{5}^{1} & \chi_{5}^{2} & \chi_{5}^{3} &\chi_{6}^{1}  &\chi_{6}^{2} & \chi_{6}^{3} &  \chi_{7}^{1} 
       & \\
      \begin{block}{c @{\hspace{10pt}}(ccccccccccc}
        \chi_{1} & 0 & -\nabla^{2} & 2\alpha\nabla^{2} & 0 & 0 & 0 & 0 & 0 & 0 & 0 & 0 \\
        \chi_{2} & \nabla^{2} & 0 & 0 & \nabla^{2}+w\nabla^{2}\nabla^{2}\nabla^{2} & 0 & 0 & 0 & 0 & 0 & 0 & 0 \\
        \chi_{3} &-2\alpha\nabla^{2} & 0 & 0 & -\nabla^{2} & 0 & 0 & 0 & 0 & 0 & 0 & 0  \\
        \chi_{4} & 0 & -\nabla^{2}-w\nabla^{2}\nabla^{2}\nabla^{2} &\nabla^{2} & 0 & 0 & 0 & 0 & 0 & 0 & 0 & 0 \\
        \chi_{5}^{1} & 0 & 0   & 0 & 0 & 0 & 0 & 0 & -\frac{1}{2} & 0 & 0 & 0 \\
        \chi_{5}^{2} & 0 & 0   & 0 & 0 & 0 & 0 & 0 & 0 &  -\frac{1}{2} & 0 & 0  \\ 
        \chi_{5}^{3} & 0 & 0   & 0 & 0 & 0 & 0 & 0 & 0 & 0 &  -\frac{1}{2} & 0  \\
        \chi_{6}^{1} & 0 & 0   & 0 & 0 & \frac{1}{2} & 0 & 0 & 0 & 0 & 0 & 0 \\
        \chi_{6}^{2} & 0 & 0   & 0 & 0 & 0 & \frac{1}{2} & 0 & 0 & 0 & 0 & 0  \\
        \chi_{6}^{3} & 0 & 0   & 0 & 0 & 0 & 0 & \frac{1}{2} & 0 & 0 & 0 & 0  \\
        \chi_{7}^{1} & 0 & 0   & 0 & 0 & 0 & 0 & 0 & 0 & 0 & 0 & 0 \\
        \chi_{7}^{2} & 0 & 0   & 0 & 0 & 0 & 0 & 0 & 0 & 0 & 0 & 0 \\
        \chi_{7}^{3} & 0 & 0   & 0 & 0 & 0 & 0 & 0 & 0 & 0 & 0 & 0 \\
        \chi_{8}^{1} & 0 & \partial^{1}  & 0 & 0 & 0 & 0 & 0 & 0 & 0 & 0 & -\frac{1}{2}\left(\nabla^{2}+\partial^{1}\partial^{1}\right) \\
        \chi_{8}^{2} & 0 & \partial^{2}   & 0 & 0 & 0 & 0 & 0 & 0 & 0 & 0 &  -\frac{1}{2}\partial^{1}\partial^{2}  \\
        \chi_{8}^{3} & 0 & \partial^{3}   & 0 & 0 & 0 & 0 & 0 & 0 & 0 & 0 &  -\frac{1}{2}\partial^{1}\partial^{3}  \\
          \end{block}
    \end{blockarray}
          \ \cdots \\
&\qquad \cdots \ 
 \begin{blockarray}{cccccc}
       \chi_{7}^{2} & \chi_{7}^{3} & \chi_{8}^{1} &\chi_{8}^{2} & \chi_{8}^{3} &\\
      \begin{block}{cccccc)}
          0 & 0 & 0 & 0 & 0\\
          0 & 0 & -\partial^{1} &  -\partial^{2} &  -\partial^{3} \\
          0 & 0 & 0 & 0 & 0 \\
          0 & 0 & 0 & 0 & 0  \\
          0 & 0 & 0 & 0 & 0\\
          0 & 0 & 0 & 0 & 0 \\ 
          0 & 0 & 0 & 0 & 0 \\
          0 & 0 & 0 & 0 & 0\\
          0 & 0 & 0 & 0 & 0 \\
          0 & 0 & 0 & 0 & 0 \\
          0 & 0 & \frac{1}{2}\left(\nabla^{2}+\partial^{1}\partial^{1}\right) & \frac{1}{2}\partial^{1}\partial^{2} & \frac{1}{2}\partial^{1}\partial^{3} \\
          0 & 0 & \frac{1}{2}\partial^{2}\partial^{1} & \frac{1}{2}\left(\nabla^{2}+\partial^{2}\partial^{2}\right) & \frac{1}{2}\partial^{2}\partial^{3} \\
          0 & 0 & \frac{1}{2}\partial^{3}\partial^{1} & \frac{1}{2}\partial^{3}\partial^{2} & \frac{1}{2}\left(\nabla^{2}+\partial^{3}\partial^{3}\right) \\
          -\frac{1}{2}\partial^{2}\partial^{1} & -\frac{1}{2}\partial^{3}\partial^{1} & 0 & 0 & 0 \\
          -\frac{1}{2}\left(\nabla^{2}+\partial^{2}\partial^{2}\right) & -\frac{1}{2}\partial^{3}\partial^{2} & 0 & 0 & 0 \\
          -\frac{1}{2}\partial^{2}\partial^{3} & -\frac{1}{2}\left(\nabla^{2}+\partial^{3}\partial^{3}\right) & 0 & 0 & 0 \\
          \end{block}
    \end{blockarray}\hspace{2mm}\delta^{3}(x-y)
 \end{aligned}
\end{equation}
and after long calculations,  it's inverse takes the form

\begin{equation}
 \begin{aligned}[b]
   \mathbf{\alpha} &= 
   \begin{blockarray}{cccccccccccc}
       & \chi_{1} & \chi_{2} & \chi_{3}& \chi_{4} & \chi_{5}^{1} & \chi_{5}^{2} & \chi_{5}^{3} &\chi_{6}^{1}  &\chi_{6}^{2} & \chi_{6}^{3} & \chi_{7}^{1} 
       & \\
      \begin{block}{c@{\hspace{10pt}}(ccccccccccc}
        \chi_{1} & 0 &-\nabla^{2}   & -\nabla^{2}-w\nabla^{6} & 0 & 0 & 0 & 0 & 0 & 0 & 0 & -\partial^{1} \\
        \chi_{2} & \nabla^{2} & 0   & 0 & -2\alpha\nabla^{2} & 0 & 0 & 0 & 0 & 0 & 0 & 0 \\
        \chi_{3} & \nabla^{2}+w\nabla^{6} & 0   & 0 & -\nabla^{2} & 0 & 0 & 0 & 0 & 0 & 0 & 0 \\
        \chi_{4} & 0 & 2\alpha\nabla^{2}   & \nabla^{2} & 0 & 0 & 0 & 0 & 0 & 0 & 0 & 2\alpha\partial^{1}   \\
        \chi_{5}^{1} & 0 & 0   & 0 & 0 & 0 & 0 & 0 & 2\Gamma & 0 & 0 & 0 \\
        \chi_{5}^{2} & 0 & 0   & 0 & 0 & 0 & 0 & 0 & 0 & 2\Gamma & 0 & 0 \\ 
        \chi_{5}^{3} & 0 & 0   & 0 & 0 & 0 & 0 & 0 & 0 & 0 & 2\Gamma & 0  \\
        \chi_{6}^{1} & 0 & 0   & 0 & 0 & -2\Gamma & 0 & 0 & 0 & 0 & 0 & 0 \\
        \chi_{6}^{2} & 0 & 0   & 0 & 0 & 0 & -2\Gamma & 0 & 0 & 0 & 0 & 0  \\
        \chi_{6}^{3} & 0 & 0   & 0 & 0 & 0 & 0 & -2\Gamma & 0 & 0 & 0 & 0  \\
        \chi_{7}^{1} & \partial^{1} & 0   & 0 & -2\alpha\partial^{1}  & 0 & 0 & 0 & 0 & 0 & 0 & 0  \\
        \chi_{7}^{2} & \partial^{2} & 0   & 0 & -2\alpha\partial^{2} & 0 & 0 & 0 & 0 & 0 & 0 & 0  \\
        \chi_{7}^{3} & \partial^{3} & 0   & 0 & -2\alpha\partial^{3} & 0 & 0 & 0 & 0 & 0 & 0 & 0  \\
        \chi_{8}^{1} & 0 & 0   & 0 & 0 & 0 & 0 & 0 & 0 & 0 & 0 & \beta(2\nabla^{2}-\partial^{1}\partial^{1})  \\
        \chi_{8}^{2} & 0 & 0   & 0 & 0 & 0 & 0 & 0 & 0 & 0 & 0 & -\beta\partial^{1}\partial^{2}  \\
        \chi_{8}^{3} & 0 & 0   & 0 & 0 & 0 & 0 & 0 & 0 & 0 & 0 & -\beta\partial^{1}\partial^{3}  \\
          \end{block}
    \end{blockarray}
          \ \cdots \\
&\qquad \cdots \ 
 \begin{blockarray}{cccccc}
       \chi_{7}^{2} & \chi_{7}^{3} & \chi_{8}^{1} &\chi_{8}^{2} & \chi_{8}^{3} &\\
      \begin{block}{cccccc)}
          0 & 0 & 0 & 0 & 0\\
          0 & 0 & -\partial^{1} &  -\partial^{2} &  -\partial^{3} \\
          0 & 0 & 0 & 0 & 0 \\
          0 & 0 & 0 & 0 & 0  \\
          0 & 0 & 0 & 0 & 0\\
          0 & 0 & 0 & 0 & 0 \\ 
          0 & 0 & 0 & 0 & 0 \\
          0 & 0 & 0 & 0 & 0\\
          0 & 0 & 0 & 0 & 0 \\
          0 & 0 & 0 & 0 & 0 \\
          0 & 0 & \frac{1}{2}\left(\nabla^{2}+\partial^{1}\partial^{1}\right) & \frac{1}{2}\partial^{1}\partial^{2} & \frac{1}{2}\partial^{1}\partial^{3} \\
          0 & 0 & \frac{1}{2}\partial^{2}\partial^{1} & \frac{1}{2}\left(\nabla^{2}+\partial^{2}\partial^{2}\right) & \frac{1}{2}\partial^{2}\partial^{3} \\
          0 & 0 & \frac{1}{2}\partial^{3}\partial^{1} & \frac{1}{2}\partial^{3}\partial^{2} & \frac{1}{2}\left(\nabla^{2}+\partial^{3}\partial^{3}\right) \\
          -\frac{1}{2}\partial^{2}\partial^{1} & -\frac{1}{2}\partial^{3}\partial^{1} & 0 & 0 & 0 \\
          -\frac{1}{2}\left(\nabla^{2}+\partial^{2}\partial^{2}\right) & -\frac{1}{2}\partial^{3}\partial^{2} & 0 & 0 & 0 \\
          -\frac{1}{2}\partial^{2}\partial^{3} & -\frac{1}{2}\left(\nabla^{2}+\partial^{3}\partial^{3}\right) & 0 & 0 & 0 \\
          \end{block}
    \end{blockarray}\hspace{2mm}\frac{1}{\Gamma}\delta^{3}(x-y),
 \end{aligned}
\end{equation}
where $\Gamma=\left(-1+2\alpha\left(1+w\nabla^{4}\right)\right)\nabla^{4}$ and $\beta=-1+2\alpha\left(1+w\nabla^{4}\right)$. Thus, the final  Dirac's  brackets are given by
\begin{equation}
\begin{split}
     \left\lbrace h_{ij},\pi^{lm}\right\rbrace_{D}=&\frac{1}{2}\left(\delta_{i}^{l}\delta_{j}^{m}+\delta_{i}^{m}\delta_{j}^{l}\right)\delta^{3}(x-y)
     \\&+\frac{1}{2\Gamma}\left(\partial_{i}\partial_{j}-\delta_{ij}\nabla^{2}\right)\left(\partial^{l}\partial^{m}-\delta^{lm}\nabla^{2}\right)\left(1-2\alpha\left(1+3w\nabla^{4}\right)\right)\delta^{3}(x-y)\\&-\frac{1}{2\nabla^{4}}\left(\left(\partial_{i}\partial^{l}\delta_{j}^{m}+\partial_{i}\partial^{m}\delta_{j}^{l}+\partial_{j}\partial^{l}\delta_{i}^{m}+\partial_{j}\partial^{m}\delta_{i}^{l}\right)\nabla^{2}-2\partial_{i}\partial_{j}\partial^{l}\partial^{m}\right)\delta^{3}(x-y),
\end{split}
\end{equation}
In the IR limit, say  $w\rightarrow 0$,  these  brackets are reduced  to those of linearized $GR$ (\ref{secondfull}). If $w\neq 0$ and $\alpha=\frac{1}{2}$, then these brackets are $w$ independent.

\section{Conclusions}
The Hamiltonian analysis for the extended $\lambda R$ model and the Ho\v{r}ava theory with smooth breaking of conformal symmetry for any value of $\lambda$ were reported. The constraints and the fundamental Dirac's brackets were found for the extended model at $\lambda\neq \frac{1}{3}$. We observed that the theory propagates three degrees of freedom at this value, one more than linearized $GR$. In addition, the $BPS$  constant can not be zero, indicating a strong coupling of the field $h_{00}$. Furthermore, at the critical point, the constraints were found and classified, then we observed that the model propagates two degrees of freedom, just like linearized $GR$. The fundamental Dirac's brackets were constructed, and we obtained those reported for $\lambda R$ gravity and linearized $GR$; the $BPS$ constant does not appear in the fundamental brackets.  \\
On the other hand, at the value $\lambda\neq \frac{1}{3}$,  the $\lambda R$-Cotton-square action presents three degrees of freedom, and the canonical structure was similar to that for the extended model at the same value. However, at the critical point,  the  action shares a canonical structure with linearized $GR$. In fact, the theory propagates two degrees of freedom, if we take the limit in the IR, say $w\rightarrow 0$, then  the fundamental brackets are reduced to those reported for linearized $GR$ and $\lambda R$ gravity. In addition, the $BSP$ constant is not restricted; if $\alpha=\frac{1}{2}$, then the brackets are  $w$ independent, this is a difference between the extended $\lambda R$ and $\lambda R$-Cotton-square gravity at the critical point. Thus, at the critical point, this  model could be a good laboratory for testing classical and quantum implications. \\
It is worth commenting that our approach introducing the variable $K_{ij}$ allows us to analyze the theories economically. Identifying the constraints was direct and very convenient for developing the canonical analysis. Also, our approach allowed us to construct the fundamental Dirac brackets that are not reported in the literature. In this manner, our results extend and complete those reported in previous works.

%----------------------------------------------------------------------------------------

%----------------------------------------------------------------------------------------

%----------------------------------------------------------------------------------------


\begin{thebibliography}{99}
\bibitem{H1}P. Hořava, Phys. Rev. D 79 (2009) 084008.
\bibitem{H2} P. Hořava, J. High Energy Phys. 0903 (2009) 020.
\bibitem{He} Herrero-Valea, M. Eur. Phys. J. Plus 138, 968 (2023).

\bibitem{OD} M. Ostrogradsky, Mem. Ac. St. Petersbourg V14, 385, (1850).
\bibitem{Stell} K. S. Stelle, Phys. Rev. D 16 (1977) 953.
\bibitem{G} Hooft, G. and Veltman, M.J.G., Ann. Inst. H. Poincare Phys. Theor. 1974, A20, 69.
\bibitem{Ba} A. O. Barvinsky, D. Blas, M. Herrero-Valea, S. M. Sibiryakov, and C. F. Steinwachs, Phys. Rev. D 93, 064022 (2016).
\bibitem{bellorin6} J. Bellorín, C. Bórquez and B. Droguett, Phys. Rev. D 108 044035 (2023).
\bibitem{bellorin7} J. Bellorín, C. Bórquez, and B. Droguett. Phys.
Rev. D 106, 044055 (2022).
\bibitem{bellorin8} J. Bellorín, C. Bórquez and B. Droguett, Phys. Rev. D 109 084007 (2024).
\bibitem{Renor}  J. Bellorín, C. Bórquez and B. Droguett, arXiv: 2405.04708.
\bibitem{blasf} Blas, D.; Pujolas, O.; Sibiryakov, S. Phys. Rev. Lett. 2010, 104, 181302.
\bibitem{bellorin2} J. Bellorín, A. Restuccia, and A. Sotomayor, Phys. Rev. D
87, 084020 (2013).
\bibitem{Dy} J. Bellorín and B. Droguett, Phys. Rev. D 98, 086008 (2018).
\bibitem{bellorin}  J. Bellorin, A. Restuccia, Int. J. Mod. Phys. D 21, 1250029 (2012).
\bibitem{Oca}  A. Escalante, P.F. Ocaña-García, Annals Phys. 465,(2024) 169662 .
\bibitem{Dirac} Dirac, P. Lectures on Quantum Mechanics. Dover Publications, Inc., New York (2001).
\bibitem{Henn} Henneaux, Marc, and Claudio Teitelboim. Quantization of Gauge Systems. Princeton University Press, 1992.
\bibitem{Bar} J. Barcelos-Neto, T.G. Dargam, Z. Phys. C 67, 701 (1995)
\bibitem{c1} C. Charmousis, G. Niz, A. Padilla, and P.M. Saffin, J. High Energy Phys. 08 (2009) 070.
\bibitem{Noemi} N. Frusciante and M. Benetti, Phys. Rev. D 103, 104060 (2021).
\bibitem{ParkD} M. Park, JHEP 09, 123 (2009).
\bibitem{ParkD2} E. Di Valentino, N.A. Nilsson, M.-I. Park, Mon. Not. R. Astron. Soc. 519, 5043 (2023).
\bibitem{ADM} R. Arnowitt, S. Deser, and C. W. Misner, in Gravitation: An Introduction to Current Research, edited by L. Witten (Wiley,
New York 1962), p. 227.
\bibitem{Fh} H. Fuhri, S. Hortner, Phys. Rev. D 103, 105014, (2021).
\bibitem{OC}  A. Escalante, P.F. Ocaña-García,  Eur. Phys. J. C 83, 1034 (2023).
\bibitem{fierz}  M. Fierz and W. Pauli, Proc. Roy. Soc. Lond. A 173, 211 (1939)
\bibitem{Hen2} M. Henneaux, A. Kleinschmidt, and G. L. Gomez, Phys.Rev. D 81, 064002 (2010).
\bibitem{bellorin3}  J. Bellorin and A. Restuccia, Phys. Rev. D 84, 104037 (2011)
\bibitem{Vic}  A. Escalante, V. A. Zavala-Perez, Can. J. Phys. 101 (2023) 
\bibitem{gh} S. Das and S. Ghosh, Mod. Phys. Lett. A, 26:2793-2801, (2011).






\end{thebibliography}
\end{document}